\documentclass{mn2e}
\usepackage{epsfig}

\title[The Dynamics of Small Star-Forming Clusters]{Accretion and
Dynamical Interactions in Small-N Star-Forming Clusters: I. N=5 case}

\author[Delgado-Donate, Clarke \&
Bate]{E. J. Delgado-Donate$^1$\thanks{E-mail: edelgado@ast.cam.ac.uk}, 
C. J. Clarke$^1$ and M. R. Bate$^2$\\
$^1$Institute of Astronomy, University of Cambridge, Madingley Road,
Cambridge, CB3 0DS\\
$^2$School of Physics, University of Exeter, Stocker Road, Exeter EX4 4QL}

\begin{document}

\date{}

\pagerange{\pageref{firstpage}--\pageref{lastpage}} \pubyear{2002}

\maketitle

\label{firstpage}

\begin{abstract}
We present results from high-resolution hydrodynamical simulations
which explore the effects of small scale clustering in star-forming
regions. A large ensemble of small-N clusters with 5 stellar {\it
seeds} have been modelled and the resulting properties of stars and
brown dwarfs statistically derived and compared with observational
data.

Close dynamical interactions between the protostars and competitive
accretion driven by the cloud collapse are shown to produce a
distribution of final masses which is bimodal, with most of the mass
residing in the binary components. When convolved with a suitable core
mass function, the final distribution of masses resembles the observed
IMF, both in the stellar and sub-stellar regime. Binaries and single
stars are found to constitute two kinematically distinct populations,
with about half of the singles attaining velocities $\geq$ 2 km
s$^{-1}$, which might deprive low mass star-forming regions of their
lightest members in a few crossing times. The eccentricity
distribution of binaries and multiples is found to follow a
distribution similar to that of observed long period (uncircularized)
binaries.
 
The results obtained support a mechanism in which a significant
fraction of brown dwarfs form under similar circumstances as those of
{\it normal} stars but are ejected from the common envelope of
unstable multiple systems before their masses exceed the hydrogen
burning limit.  We predict that many close binary stars should have
wide brown dwarfs companions. Brown dwarfs and, in general, very low
mass stars, would be rare as {\it pure} binary companions. The binary
fraction should be a decreasing function of primary mass, with
low-mass or sub-stellar primaries being scarce. Where such binaries
exist, they are either expected to be close enough (semi-major axis
$\sim$ 10 AU) to survive strong interactions with more massive
binaries or else born in very small molecular cloud cores.
\end{abstract}


\section{Introduction}

The advent of infrared imaging has in recent years brought about the
realization that virtually {\it all} stars are formed in clusters
(e.g. Clarke et al. 2000). On a smaller scale, most field stars are
known to be binaries (e.g. Duquennoy \& Mayor 1991) or higher order
multiples (e.g. Tokovinin \& Smekhov 2002), and there is convincing
evidence that the multiplicity among young stars is even larger
(Reipurth \& Zinnecker 1993; Ghez et al. 1993; K\"{o}hler \& Leinert
1998). Even at earlier stages, Reipurth (2000) reports on a binary
frequency as high as $\sim$~80~\% for sources driving giant
Herbig-Haro flows. Additionally, the evidences for coevality of young
binary systems (e.g. White \& Ghez 2001) further suggest that
clustering at the scales of molecular cloud cores must be a common
property of star-forming regions. As shown by Sterzik \& Durisen
(1998), dynamical processes occurring within these small stellar
groups at birth are likely to affect the properties of the resulting
stellar objects. Similarly, gas dynamics are equally important at this
stage: circumstellar discs are likely to influence the dynamics of
close encounters (McDonald \& Clarke 1995) and differential accretion
from a common envelope can imprint a large range of dynamical masses
on the stars involved (Bonnell et al. 1997).

Additionally, from a theoretical point of view, the fragmentation of a
cloud core into a small aggregate of stars can help to attenuate the
{\it angular momentum problem} in star formation (Mestel \& Spitzer
1956; Bodenheimer 1995). As suggested by Larson (2002), the formation
of a binary or multiple system can redistribute the excess angular
momentum of the infalling envelope into the orbital motion of the
stellar companions. Other mechanisms such as transport processes in
discs and outflows might not be as efficient.

An often used argument against the paradigm of small N clustering at
birth is based on the sub-millimetre results of Motte et al. (1998),
who showed that the mass function of dense cores in Ophiucus is very
similar to the stellar IMF. This important result has been widely
interpreted as favouring a one to one mapping from core to star,
leaving rather limited role for multiple star formation. Since direct
evidence for such substructure is hard to address observationally
(small N clusters disintegrate over a few crossing times), it is
interesting to explore the consequences that this hypothesis can have
on the initial mass function and the orbital parameters of binary
stars, among other observables. The present paper will focus on small
aggregates of N=5 protostars set in an initially uniform background of
molecular gas, where the potential of the cluster is initially
dominated by the gaseous component. Accretion and dynamical
interactions take place naturally in this scenario and provide a
diversity of mechanisms by means of which close binary stars, multiple
systems and brown dwarfs can be formed, and their properties
statistically compared with observational data.

The structure of the paper is as follows. In Section~2 
the progress in our understanding of the fragmentation of gas
clouds and the disintegration of multiple systems is reviewed, Section~3
describes the calculations performed, with a caveat on the
role of circumstellar discs. Section~4 gives a detailed
account of the evolution of the clusters. In Section~5 the results
{\it per given core mass} are presented. Section~6 introduces the
convolution with a core mass function. The properties of the brown
dwarfs formed are discussed in Section~7. Finally, our
conclusions are given in Section~8.

\section{Previous Work}

\subsection{The disintegration of small N clusters}

The earliest studies of the disintegration of small clusters treated
only point mass dynamical interactions (van Albada
1968a,b). Subsequently, a number of authors have investigated the pure
N-body problem using a variety of numerical codes (e.g. Harrington
1974, 1975; Mikkola and Valtonen 1986). In general, the outcome of
such simulations is the break up of the cluster after $\sim$~N
crossing times, the energy for the break up being released by the
dynamical formation of a bound subsystem, usually a binary star
containing the two most massive cluster members.  As pointed out by
McDonald \& Clarke (1993), such an outcome has immediate consequences
for binary pairing statistics, implying a predilection for binary
membership amongst massive stars and a {\it dynamical bias} against
brown dwarfs as binary companions.  Another obvious implication of
this formation mode is that the kinematics of young stars are likely
to reflect the hardness of the three body encounters that ejected them
from the cluster: for a cluster containing stars of typical mass M
which suffer such an encounter at radius $r_{close}$, the typical
ejection velocity is $\sim (GM/r_{\rm close})^{1/2}
\sim 1.5 r_{c100}^{-1/2} {\rm km s}^{-1}$, where $r_{c100}$ is
normalised to an encounter at $100$ A.U and a mass of
1M$_\odot$. Sterzik \& Durisen (1995) were the first authors to link
ejections from compact clusters to the generation of Runaway T Tauri
stars (RATTs), whilst Armitage \& Clarke (1997) pointed out that the
ejection process would be likely to prune the circumstellar discs of
such stars and thus render them less likely to be Classical T Tauri
stars.  To date, the most accurate and statistically complete study of
the pure N-body disintegration problem is contained in the series of
papers by Sterzik \& Durisen and collaborators (Sterzik \& Durisen
1995, 1998; Durisen et al. 2001), which analyse the decay channels and
resulting kinematics and binary properties of a large ensemble of
Nbody simulations. This set of simulations provides an invaluable
benchmark against which to compare hydrodynamical simulations.

It is evident that in the case of young stars with discs, encounters
at separations less than about 100~AU will not in any case be
dominated by point mass gravity, but will be further modified by
dissipative interactions with circumstellar discs. Clarke \& Pringle
(1991) and McDonald \& Clarke (1995) modelled the effect of discs by
adding parameterised drag terms during close encounters and found that
the possibility of energy loss via the disc weakened the dynamical
bias in binary pairing characteristics - in other words, lower mass
stars ended up in binaries with a higher probability, especially as
binary secondaries, compared with the results of pure Nbody
simulations.

In all the above calculations, the mass of each star is assigned at
the outset, so that although they yield predictions for binarity and
stellar kinematics, they can say nothing about the form of the
IMF. Such pre-assignment of masses corresponds to the case that most
of the mass of the parent core is contained in non-linear density
perturbations at the onset of gravitational instability.  The opposite
extreme would be a model in which gravitationally unstable
fluctuations contain initially only a small fraction of the core's
mass, with the remainder being distributed throughout the core. In
this latter case, the gravitationally unstable {\it seeds} acquire
most of their mass through subsequent accretion, which is competitive
in the sense that all {\it seeds} attempt to {\it feed} from the same
reservoir. In such calculations, the final masses of stars are
determined by the same dynamical processes that create the binary
pairs and lead to stellar ejections. Such ideas were first explored in
the simulations of Bonnell et al. (1997), who showed that due to the
inequitable nature of the competitive accretion process, a large
dynamic range of final stellar masses will result even in the case
that the seeds all start off with equal masses. More recently,
Reipurth \& Clarke (2001) have specifically linked the cluster
formation scenario to the production of brown dwarfs, arguing that the
low mass end of the mass spectrum was populated by {\it seeds} that
had lost out in the competitive accretion process and were prematurely
ejected.

\subsection{Hydrodynamics: The fragmentation of molecular cloud cores}

The favoured mechanism for the formation of most binary stellar
systems is the fragmentation of a collapsing molecular
cloud. Fragmentation has been studied numerically for $\sim$ 20
years (Boss \& Bodenheimer 1979; Boss 1986; Bonnell et al 1991; Nelson
\& Papaloizou 1993; Bonnell 1994; Burkert \& Bodenheimer 1993; Bate,
Bonnell \& Price 1995; Truelove et al. 1998). These calculations
appear to show that it is possible to form binaries with similar
properties to those that are observed in fragmentation
calculations. However, these simulations lack predictive power since
the results depend sensitively on the initial conditions, which are
poorly constrained. Besides, following the calculation significantly
beyond the point at which fragmentation occurs is extremely
computationally expensive. Bate \& Bonnell (1997) quantified how the
properties of a binary system are affected by the accretion of a small
amount of gas from an infalling gaseous envelope. They found that the
effects depend primarily on the specific angular momentum of the gas
and the binary's mass ratio. Generally, accretion of gas with low
specific angular momentum decreases the mass ratio and separation of
the binary, while accretion of gas with high specific angular momentum
increases the separation and drives the mass ratio towards unity (see
also Artymowicz 1983, Bate 2000). From these results, they predicted
that closer binaries should have mass ratios that are biased toward
equal masses compared to wider systems, since the gas which falls on
to a closer system is likely to have more specific angular momentum,
relative to the binary's, than for a wider one.

A more direct approach has recently been conducted by Bate et
al. (2002a,b), in which a gas cloud which is subject to a supersonic
turbulent velocity field is modelled. The divergence-free velocity
field rapidly generates a richly non-linear density structure in the
gas (Ostriker, Stone \& Gammie 2001). Dense cores are continually
created and destroyed and, eventually, some of them become Jeans
unstable and collapse (e.g. see Padoan \& Nordlund 2002 for
details). However, these cores possess internal sub-structure and
fragment further until the opacity limit for fragmentation (a few
Jupiter masses) is reached. Thereafter, the stellar {\it seeds} grow
in mass by accretion and interact strongly with other collapsed
fragments and protostellar discs in their vicinity. This system
readily demonstrates the formation of small N ensembles in which the
sort of behavior described in the last section is observed to
occur. Currently, however, the accuracy of the statistical properties
of the stars that can be determined from such a one-off calculation
are limited by the small number of objects ($\approx$~50)
formed. Additionally, the computational expense involved in this
large-scale simulations makes it prohibitive to explore different
initial conditions. Modeling the evolution of many small-N ensembles
allows us to obtain more accurate statistics and to investigate the
dependence of the results on the initial conditions imposed.

\section{The simulations}

The calculations reported in this paper have been performed using a
three-dimensional smoothed particle hydrodynamics (SPH) code based on
a version originally developed by Benz (Benz 1990; Benz et
al. 1990). The smoothing lengths of particles vary in time and space,
such that the number of neighbours for each particle remains
approximately constant at N$_{\rm neigh}$~=~50. We use the standard form
of artificial viscosity (Monaghan \& Gingold 1983) with strength
parameters $\alpha_v$~=~1 and $\beta_v$~=~2.

The stellar {\it seeds} are modelled by sink particles (Bate, Bonnell
\& Price 1995), which interact only via gravity. Sink particles are
characterized by a constant sink radius R$_{\rm sink}$ such that any
gas matter which falls into it and is bound to the sink particle is
accreted. The gravitational acceleration between two sink particles is
Newtonian for r~$\geq$~R$_{\rm sink}$, but is smoothed within this
radius using spline softening (Benz 1990). The maximum acceleration
occurs at r~$\sim$~R$_{\rm sink}$/4; therefore this is the minimum
binary separation R$_{\rm min}$ that can be resolved. The sink radius
has been chosen to be $10^{-3}$ of the initial radius of the cloud, so
that R$_{\rm min}$ is $\sim$~1~AU.  The gas component is modelled
using $10^4$ SPH particles with equal masses.

\subsection{Initial conditions}

One hundred simulations of these small N ensembles has been performed.
Each simulation consists of an isothermal spherical gas cloud, with
initial constant density. Five stellar {\it seeds} (sink particles)
were randomly placed inside a sphere of radius R$_*$ equal to one
fourth of the initial cloud radius R$_{\rm cl}$. The choice of N=5 is
motivated by two reasons: first, we want to study the effect of
dynamical interactions in unstable multiple systems (therefore N must
be greater than 2) and second, we would like to allow for the
possibility of binary-binary encounters to take place (thus, at least
N=4). The precise choice of N=5, rather than 4 or 6 is not
particularly relevant, since the results are not sensitive to small
variations in the number of {\it seeds}. Results from clusters with a
larger number of seeds, say N=10,20, will be provided in a future
paper. Regarding the initial density distribution, the choice of a
homogeneous density profile has been made following the properties of
observed cores: e.g. Alves et al. (2001) found that the Bok globule
B68 could be fitted remarkably well by a Bonnor-Ebert distribution,
which is characterised by a flat inner profile; many other
observations support this same result (see Andr\'e et al. 2000 for a
review). In addition, theoretical work by Bate (2000) has shown that
cloud cores with initially constant density profiles are more likely
to produce binary systems with properties more similar to those that
are observed. A steeper initial density profile would enhance
dramatically the effects of competitive accretion, consecuently
producing stellar objects with too unequal masses (see Section~4 for
details on how competitive accretion works.)

The initial mass of the cloud M$_{\rm cl}$ is a scale-free quantity
and therefore is set to 1~M$_\odot$ by default, so that the final
stellar mass gives directly the cloud mass fraction that was accreted
by that object. Obviously, the final {\it real} masses will be given
by the product of that mass fraction times the mass of the core, taken
randomly from a certain core mass probability distribution.

The gas component comprises 90~\% of the initial mass, and is
initially in virial equilibrium with its own gravitational
potential. This condition determines the radius of the cloud to be
equal to:
\begin{equation}
R_{\rm cl} = \frac{2}{5} G M_g \frac{\mu}{R_g T}   
\end{equation}
where $G$ is the gravitational constant, $\mu$ is the mean molecular
weight (the gas is assumed to consist of pure molecular hydrogen and
therefore $\mu$ is set to 2), $R_g$ is the gas constant, $T$ is the
temperature, and $M_g$ is the gas mass.

Correspondingly, the {\it protostars} are also set in virial
equilibrium with the gas. This condition determines the initial
velocity dispersion of the seeds. Note that virial equilibrium means
that, in most simulations, if only the gravitational force between
{\it seeds} is computed, the majority of the protostars are bound to
each other, i.e., they form initially non-hierarchical high-order
multiples. The absolute value of the binding energy of the pairs,
however, is very low, of the order of several~$\times$~10$^{-5}$ times
the binding energy of the resulting tightest pair. The cluster is
bounded by a small external pressure to prevent the escape of gas
particles at the initial stages when the gas thermal energy is
comparable with its gravitational energy. The gas is initially
stationary and thus the total angular momentum is equal to zero. The
initial $\beta$ parameter (ratio of the rotational energy of the
protostars to the potential energy of the gravitational interaction
between the protostars and the gas cloud) of the stellar system has a
mean value of 0.03, i.e., the initial angular momentum of the stellar
{\it seeds} is not negligible but certainly low. Thus, we favour
models in which the stars lack rotational support against the pull of
the gas once this collapses towards the centre.

The numerical values of all the cluster fundamental parameters are
given in Table~1.

\begin{table*}
\begin{center}
\begin{tabular}{lcccccccccccc}
\hline
M$_{\rm cl}$~(M$_\odot$) & R$_{\rm cl}$~(AU) & M$_g$~(M$_\odot$) & M$_*$~(M$_\odot$)
& T~(K) & $\alpha$ & $\sigma_*$~(km s$^{-1}$) & R$_{\rm sink}$~(AU) &
R$_{\rm min}$~(AU) & d$_{\rm *,ini}$~(AU) \\ \hline
1 & 8770 & 0.9 & 0.1 & 10 & 1 & 0.2 & $\approx$~10 &
$\approx$~1 & $\approx$~10$^3$ \\ \hline
\end{tabular}
\caption{\label{setup} Initial parameters of the simulations. The
meaning of the columns is as follows $\rightarrow$ M$_{\rm cl}$ :
Total mass of the cloud; R$_{\rm cl}$: Initial radius of the gas
cloud; M$_g$: Initial amount of mass in gas; M$_*$: Initial amount of
mass in the stellar {\it seeds}; T: Temperature; $\alpha$: Ratio of
internal to gravitational energy (a value of 1 means that the cloud
contains initially one thermal Jeans mass); $\sigma_*$: Initial
velocity dispersion of the stellar {\it seeds}; R$_{\rm sink}$:
Accretion radius of the sink particles; R$_{\rm min}$: Radius of
maximum gravitational acceleration; it sets the spatial resolution of
the calculations; d$_{\rm *,ini}$: Initial mean distance between
stellar {\it seeds}}
\end{center}
\end{table*}

An {\it efficiency} of 100~\% is assumed in all calculations: gas is
allowed to be accreted until all the cloud mass is in the form of
stars. Although this is not realistic, the lack of firm constraints on
star-formation efficiency of molecular cloud cores has deterred us
from switching the gas accretion off at any arbitrary time, thus
allowing the investigation of this process at all stages of the
cluster evolution.

An unstable multiple is likely to remain in the cluster once all the
gas has been exhausted. Thereafter, the calculation is followed using
a pure N-body algorithm. This has been done for approximately 80 \% of
the simulations. The N-body algorithm used works under the same
principles as the SPH code: adaptive time-steps evolved using a
Runge-Kutta-Fehlberg integrator which provides a spatial resolution
limited by the softening radius $\epsilon$~$\sim$~R$_{\rm sink}$ . No
regularization procedure has been applied.

Each of these simulations required an average of $\approx$~450~CPU
hours on Sun UltraSPARC 1/5/10 workstations for the hydrodynamical
calculations. The N-body procedure needed an average of
$\approx$~350~CPU hours, the standard deviation of this distribution
being much larger than in the hydrodynamical case due to the
inequitable timescales involved in the random process of dynamical
decay.  To test the dependency of the results on the numerical
resolution, calculations were also performed with 5~$\times$~10$^4$
particles. The results did not show any qualitative difference from
the 10$^4$ particles run.

These simulations do not include magnetic fields nor radiative
transfer. An isothermal equation of state (T~=~10~K) is used
throughout the calculations to resemble the approximate balance between
compressional heating and radiative cooling characteristic of the
low-density regions of molecular clouds. No feedback from the stars
has been modelled. This may be an adequate choice for low-mass star
formation where strong stellar winds are not expected.

\subsection{Circumstellar discs}

Protostellar discs are an intrinsic byproduct of the star formation
process. Disc formation occurs whenever any non-zero angular momentum
is present in the initial gas from which the stars form. These discs
can be quite large and can thus affect the dynamics of the stellar
interactions (McDonald \& Clarke 1995; Hall et al. 1996). The
calculations reported here do not include any discs in the initial
conditions. The reason for this is twofold: first, to place discs at
the onset of the simulations would require a large number of SPH
particles, which would subsequently slow down significantly the
calculations and therefore make the choice of a statistical approach
virtually impossible. Secondly, the initial conditions in the gas cloud
do not provide enough vorticity to generate large protostellar discs;
initial velocity perturbations such as turbulent fields would do this
but, again, would require a computationally prohibitive
number of gas particles. Global rotational motion in the cloud is
known to generate vorticity and will be explored elsewhere.

The role of protostellar discs is to provide close encounters with an
additional dissipative term which can affect the pairing
characteristics of binary stars. Clarke \& Pringle (1991) and McDonald
\& Clarke (1995) included discs in their model 
and found that the possibility of
energy loss via disc weakened the dynamical bias in binary pairs -- in
other words, more lower mass stars ended up in binaries, especially as
binary secondaries, compared with the results of pure N-body
simulations. On the other hand, the inclusion of gaseous discs
introduces a mechanism for the formation of stellar objects via
 fragmentation induced by gravitational
instability (Bonnell 1994; Whitworth et al. 1995; Burkert, Bate \&
Bodenheimer 1997; Bate et al 2002a,b), process
that could set a natural scale for the initial mean separation of some
of the interacting fragments smaller than that assumed
here. Additionally, circumbinary discs can have an important effect on
the evolution of the binary orbital elements: the
binary may transfer orbital angular momentum to the material in the
disc via gravitational torques (Artymowicz et al. 1991; Bate \&
Bonnell 1997); this loss of
angular momentum reduces the separation of the binary.

The effects of circumstellar disc on the dynamics of these small N
clusters will be explored in the future.

\section{The evolution of Small-N clusters}

\subsection{Onset of instabilities, cloud collapse and the formation of a binary system}

As each calculation begins, the protostellar seeds start accreting gas
from the common envelope.  The initial gas density is constant and
there are no pressure gradients to oppose the cluster self-gravity,
which therefore begins to collapse. The stellar motions generate small
perturbations in the gas density, producing local departures from the
expected r$^{-2}$ profile. Such perturbations arise from the accretion
of gas that is in the vicinity of the stars: under-dense regions are
created and, in turn, replenished by the infall of gas from larger
distances. This replenishment is dependent not only on the
gravitational attraction of the individual stars but mostly on the
overall gravitational potential of the cluster.

The presence of gas in the cluster has two main effects on the
evolution of the system's energy budget (see Bonnell et
al. 1997). First, there is mass loading, where the gas is accreted on
to the individual stars, increasing their masses without changing
their momenta and hence decreasing their kinetic energies. Second, the
presence of gas acts as a drag source on the stars. The motion of the
{\it seeds} through the cluster accelerates and condenses the gas
behind the stars, which subsequently experiences isothermal shocks
that radiate the additional kinetic energy away. More importantly, the
cloud collapse proceeds isothermally and therefore the gas
gravitational energy increases enormously (in absolute value), {\it
dragging in} the stars, until enough accretion on to the seeds turns
the stellar component into the main focus of gravitational attraction.
These two processes contribute to making the cluster more bound,
increasing the chances of forming very tight binaries.

Once the collapse of the cloud drives the stars towards the cluster
centre, where the density is larger, accretion can proceed more
efficiently. The accretion rate of a star depends predominantly on its
position in the cluster. Those stars closest to the centre have the
benefit of the cluster potential in attracting gas. The stars that, at
the time of collapse, stay at the outside of the cluster never accrete
appreciable amounts of gas. This differential accretion process will
be decisive in providing a large range of final stellar masses.  As
soon as one object accretes a significant fraction of the gas
reservoir, it starts to dominate the potential of the cluster and,
therefore, turns into the main component of an unstable multiple
stellar system. Several possibilities are thereafter allowed by the
chaotic nature of these simulations.

\begin{figure}
\begin{center}
\centerline{\epsfig{file=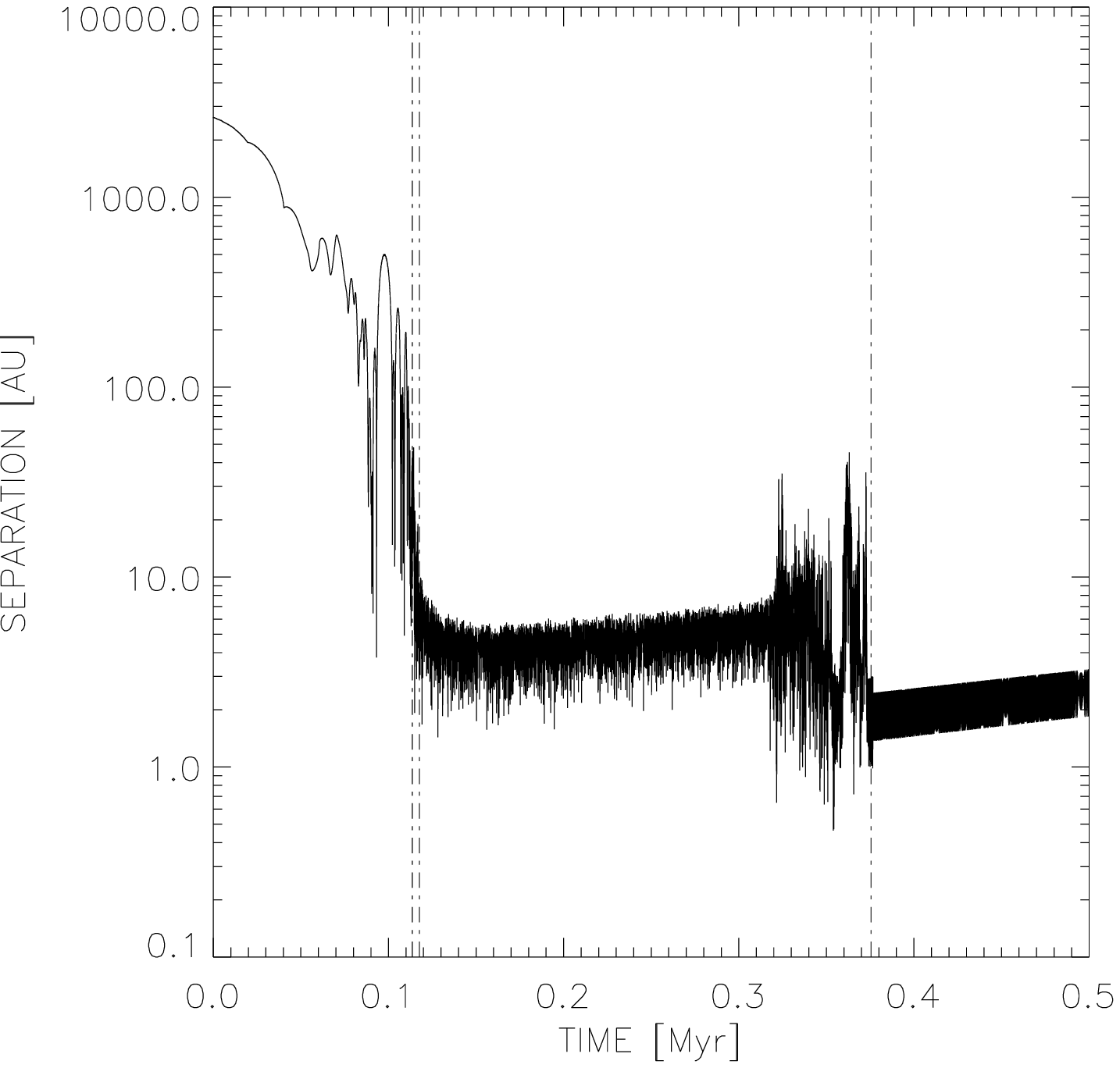,height=4.5cm}\epsfig{file=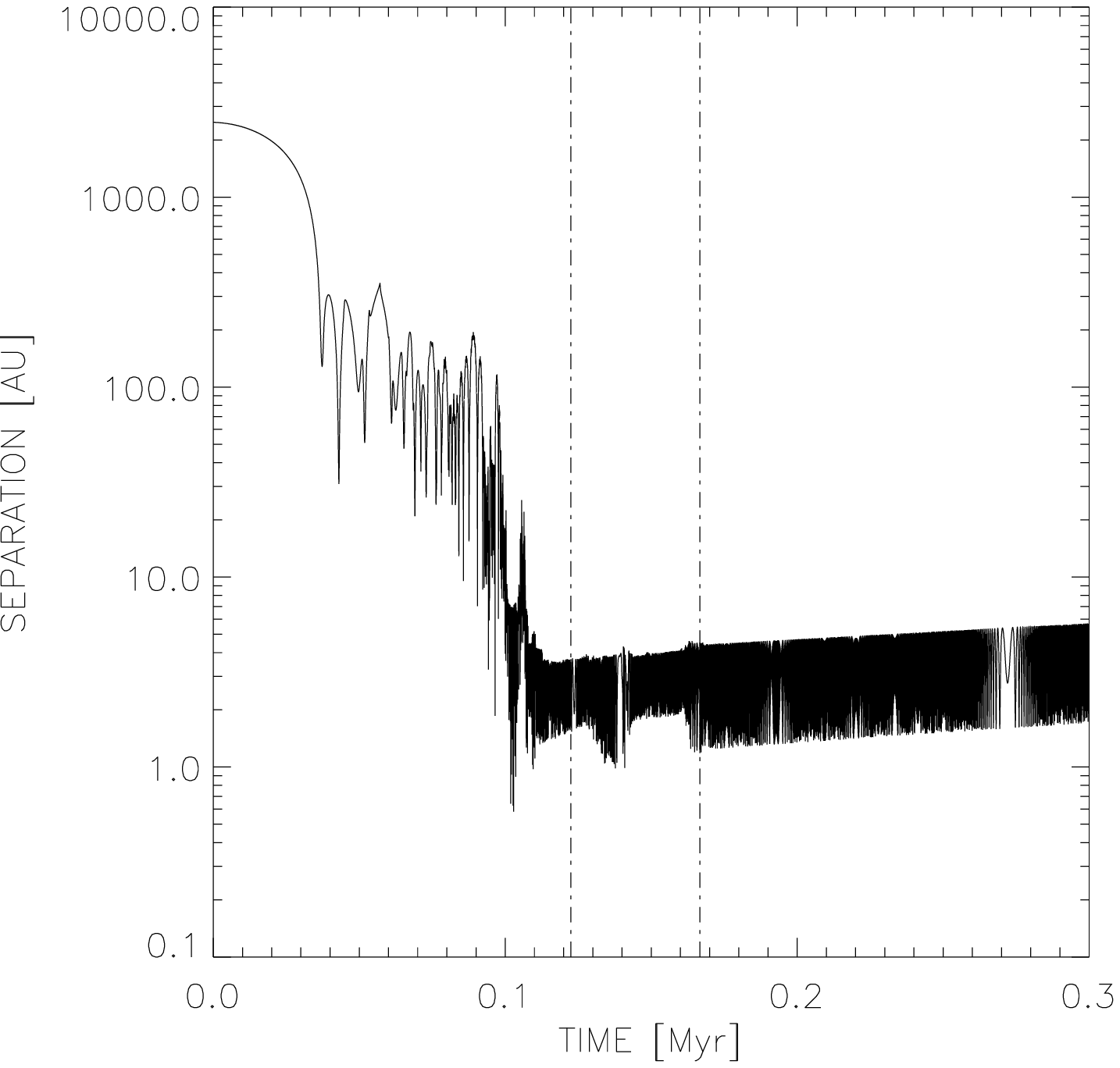,height=4.5cm}}
\centerline{\epsfig{file=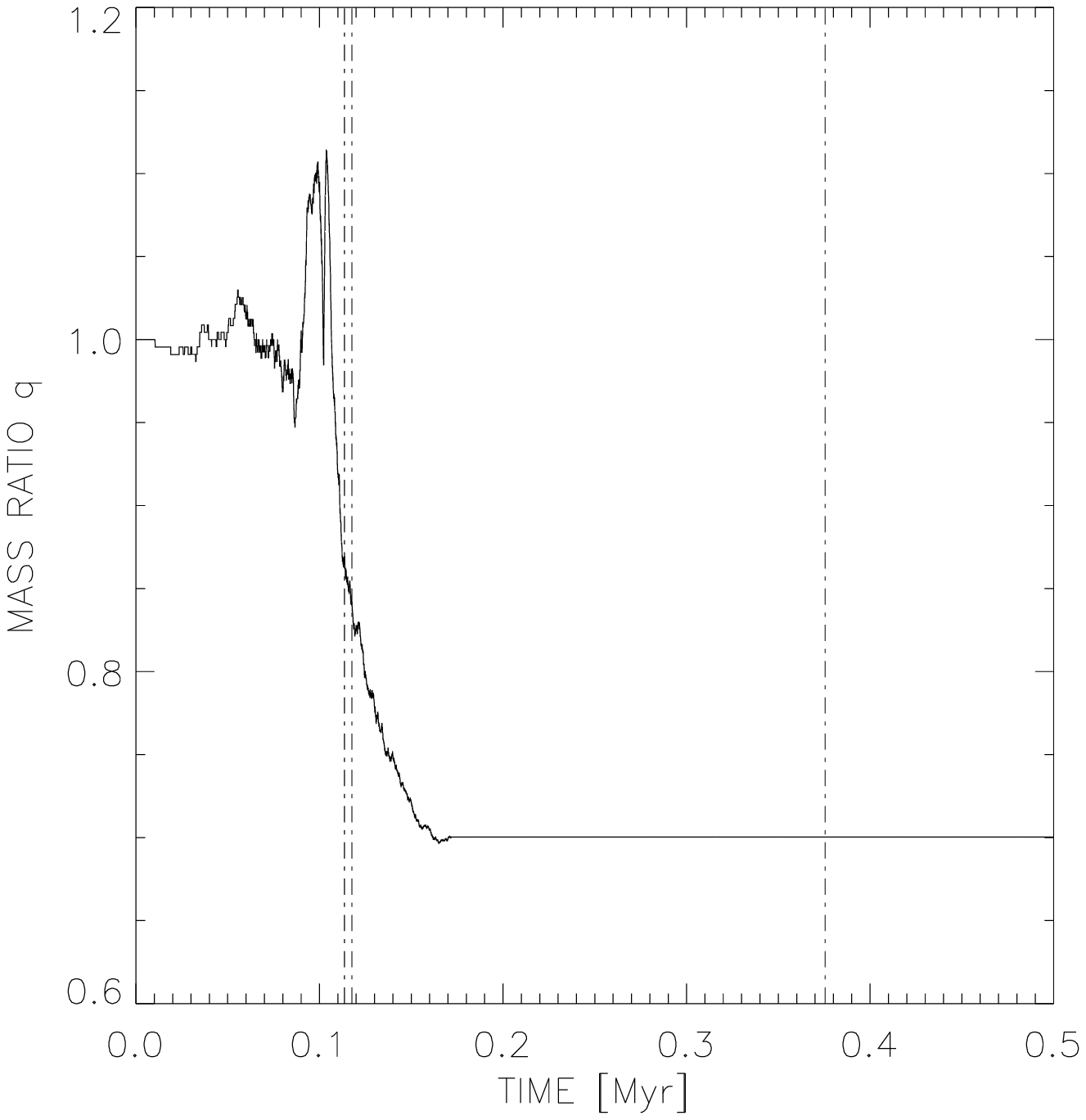,height=4.5cm}\epsfig{file=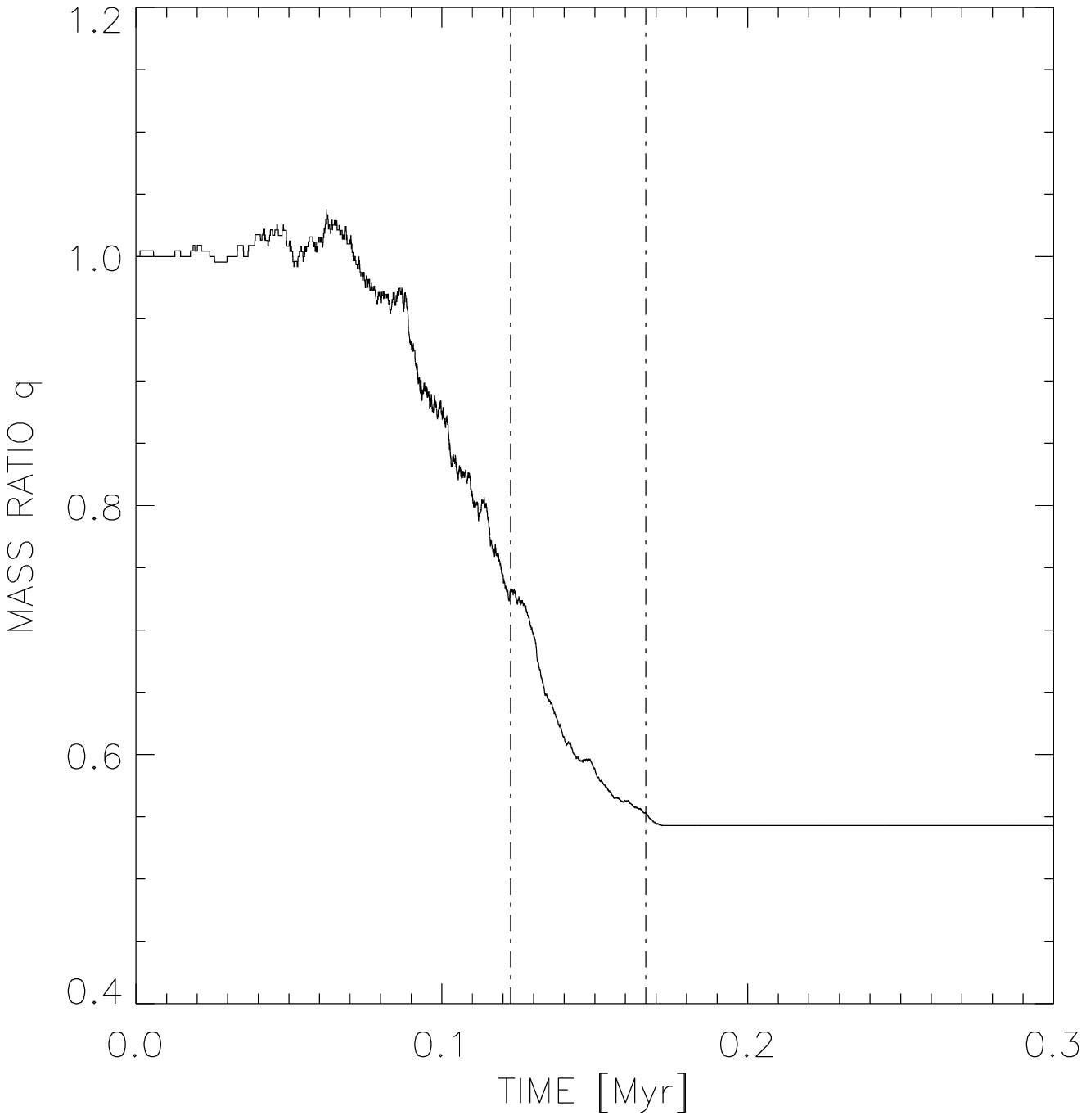,height=4.5cm}}
\caption{Top left: Time evolution of the separation and, bottom left,
the mass ratio $q$ of a binary star system. On the right: same as left
diagrams but for the inner pair of a triple system.  The time axis is
given in Myr and the separation in AU. The dot-dashed lines indicate
the ejection times of single stars. As in these two cases, the binary
mass ratio decreases monotonically in most of the simulations.  Taking
into account that most of the binaries end up with a mass ratio in the
0.5-1 range, we conclude that the values of $q$ are quite insensitive
to differential accretion and dynamical interaction processes
occurring in the cluster. See Section~4.1 for details.}
\end{center}
\end{figure}

First, two seeds may start close enough so as to soon form a binary,
therefore enhancing accretion onto the system and out-weighting the
effect of competitive accretion between the two components. At this
early stages, the {\it seed} masses are still close to their initial
values and therefore, the binary mass ratio is close to unity. Since
no initial rotation is present in the cloud, the specific angular
momentum of the infalling gas about the binary rotation axis will be
in general lower than the specific angular momentum of the
binary. This makes the binary mass ratio depart slowly from unity and
the binary semi-major axis progressively decrease (see
Section~2.2). Subsequent dynamical interactions can change
significantly the semi-major axis of the binary, produce ejection of
lower-mass {\it seeds} and break up the stellar system into a
hierarchical multiple, but not change the binary mass ratio
appreciably. The evolution of the separation and mass ratio of the two
inner components of a binary and a triple system are shown in
Fig.~1. Note how the semi-major axis of the binary changes abruptly
when an ejection takes place, whilst the mass ratio decreases slowly
due to the differential gas accretion of the two components.

Second, it may happen that only one object accretes a significant
amount of gas before any close binary system is formed. In this
situation, the other {\it seeds} are attracted towards the most
massive member and acquire very eccentric orbits. The high
eccentricity enhances the differential accretion as only at pericentre
is the density high enough to make gas accretion efficient. The result
of this process is to drive the mass ratio of the binary to small
values, although never too far from unity (binaries with mass ratios
smaller than 0.2 are rare).

Of course, most of the simulations performed display an accretion
history that lies somewhere in between the two extreme cases described
above. Yet the common outcome is that most of the cluster mass ends up
in the binary system, as it settles in the gas-rich centre very early
on in the cluster evolution, and that the mass ratio of this binary
depends substantially on how soon the pairing occurs relative to the
accretion history of the {\it seeds}.

In summary, the consequences of gas accretion are, among others, to
produce an unstable multiple system frequently dominated by a binary
star whose mass ratio has, to a significant degree, been set. From
this point of the calculation onwards (with about 70~\% of the gas
matter left) dynamical interactions play a major role.

\subsection{Dynamical interactions: ejections, binary hardening and hierarchical multiples}

Dynamical interactions become important when the cluster evolution is
close to its first free-fall time. The central binary has accreted a
significant fraction of the available gas and begins to dominate the
cluster potential. The other three {\it seeds} are attracted to its
vicinity and very close encounters soon take place. The effect of
these encounters is twofold: first, repeated three-body encounters
between the central binary and the incoming object contribute to
harden the binary, therefore decreasing abruptly its semi-major axis,
by a factor ranging from 2 to 10, and increasing its eccentricity. The
sudden growth of eccentricity can have a small long-term effect on the
mass ratio evolution of the central binary, as explained in
Section~4.1.

Secondly, the incoming object, with no exception a lower-mass member
of the cluster, is ejected. Ejection means that either this object
stabilizes into a hierarchical orbit (at a distance much larger than
the new semi major axis of the binary), or it attains an outward speed
larger than the cluster present escape velocity, therefore leaving the
cluster forever. In the latter situation, the ejection halts any
further accretion, therefore setting the final mass of the escaper and
its kinematical properties.  An object which remains bound to the
cluster in a large orbit may also be deprived of further gas accretion
if the gas is already too centrally-condensed.

In general, these simulations show that at least one object and
sometimes two are ejected and escape from the cluster at an early
stage, when about 70~\% of the gas matter is still present. These
escapers leave with a very low fraction of the cluster mass and, on
average, need only a very soft kick to escape the potential well. The
typical ejection velocities of these objects are $\approx$~1~km
s$^{-1}$.

On the other hand, it is common that a non-hierarchical triple system
survives the cluster break-up once all the gas has been exhausted. The
decay of this unstable multiple is a stochastic process that needs to
be followed numerically using pure N-body algorithms. Sometimes the
triple system disintegrates into a hardened binary and a single. The
single star, frequently more massive than those ejected at previous
times, receives also a stronger kick as the distance of close approach
can be as small as $\approx$~1~AU (i.e. as low as it is allowed by the
numerical resolution imposed). In other runs, the system stabilizes
into a hierarchical triple. To consider when this is the case, the
two-step criterion proposed by Eggleton \& Kiseleva (1995) has been
used. First, the triple system is partitioned into two pairs; this can
be done in three different ways (e.g. [1-2]-3, [1-3]-2, [2-3]-1). For
each such configuration, two {\it instantaneously} Keplerian orbits
can be computed: the {\it inner} orbit of two of the bodies in their
centre of mass (CM) frame, and the {\it outer} orbit of the third body
around the CM of the {\it inner} component in the CM frame of the
whole system. For the system to be a instantaneous hierarchical
triple, both orbits must be bound (negative energy) at least in one of
the configurations, and the ratio of the {\it outer} to {\it inner}
periods must be greater than unity . If there is more than one
configuration in which both orbits are bound, that with the largest
ratio of {\it outer} to {\it inner} periods is selected. Second, if at
a given time the triple system is found to fulfill the previous
criterion, the ratio of the {\it outer} pericentre distance to the
{\it inner} apocentre distance is computed. If this ratio is greater
than a critical value Y$_{0}^{\rm min}$:
\begin{equation}
Y_{0}^{\rm min} \approx 1 + \frac {3.7} {q_{\rm out}^{1/3}} + \frac
{2.2} {1 + q_{\rm out}^{1/3}} + \frac {1.4} {q_{\rm in}^{1/3}} \frac
{q_{\rm out}^{1/3} -1} {q_{\rm out}^{1/3} + 1}
\end{equation}
which depends on the {\it inner} $q_{\rm in}$ and {\it outer} $q_{\rm
out}$ mass ratios, then the system is said to be {\it stable} for a
time 10$^n$~$\times$~the triple {\it outer} period, where $n$ is
larger than 4 in $\sim$~95~\% of the cases. This latter criterion
guarantees that a substantial fraction of the triples included in our
results have survival times greater than 100~Myr.

Finally, approximately 15~\% of the simulations produce hierarchical
quadruple systems, the only escaper being invariably ejected at very
early times. Quadruples have been identified using a generalized
version of the criterion described in the previous paragraph.  In very
few cases (2~\%), two mutually unbound binaries are produced, plus one
single star. One of these simulations is particularly striking as,
contrary to the rule, the total mass in the two binaries is much
smaller than that of the single.

\section{Statistical results {\it per core mass}}

\subsection{Mass function}

\begin{figure}
\begin{center}
\centerline{\epsfig{file=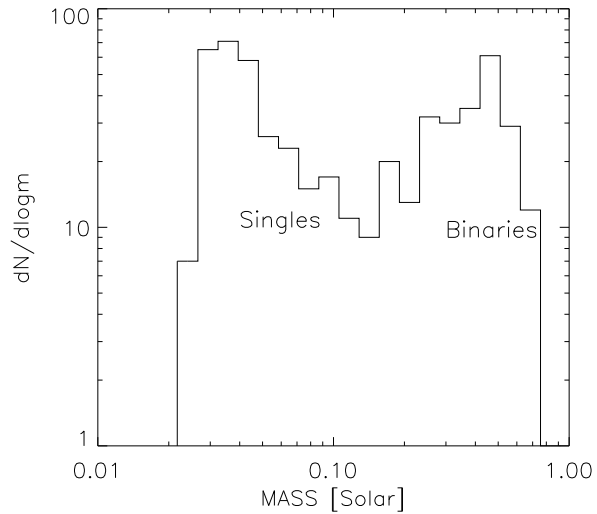,height=7.5cm}}
\caption{ Stellar Mass Fraction Probability Distribution
(SMFPD): distribution of stellar mass {\it per given core mass}, i.e.,
the final mass obtained in the simulations assuming that the initial
mass of the cores is set to 1~M$_\odot$. The bimodal shape results
from the inequitable mass accretion between single stars and outer
companions (low-mass peak), and binary stars or inner pairs of
multiple systems (high-mass) peak.}
\end{center}
\end{figure}

\begin{table}
\begin{center}
\begin{tabular}{lcccc}
\hline
f$_b$ & f$_t$ & f$_q$ & f$_m$ \\ \hline
0.154 & 0.108 & 0.045 & 0.307 \\ \hline
\end{tabular}
\caption{Fraction of binaries (f$_b$), triples (f$_t$) and quadruples
(f$_q$), out of the total number of systems (singles $+$ binaries $+$
...); f$_m$ refers to the fraction of all multiple systems
altogether.}
\end{center}
\end{table}

Table~2 shows the frequency of binary, triple and quadruple systems
found in these simulations. The frequency $f$ is defined as the
fraction of binaries/triples/quadruples over the total number of
systems, where a system can be a single, a binary, etc...(e.g. if out
of five stars, two are in a binary, the binary fraction would be
$\frac {1} {4}$). 47~\% of the simulations produce one binary star and
three single stars. The binary mass ratio is in the range 0.5-1 and
the three escapers are of much lower mass. In a minority of cases only
two or one stars (36~\% and 15~\%, respectively) are ejected as
singles, with the third (and fourth) being retained at large
separations. Once again, the binary contains most of the mass, and the
multiple companion(s) are, like the single stars, of very low
mass. This pattern of mass acquisition is reflected in the mass
distribution of stars produced for a given {\it parent core mass} (see
Figure~2). This Stellar Mass Fraction Probability Distribution
(hereafter SMFPD) is bimodal: one peak corresponding to the binary
members and the other to the low mass singles and wide companions.

\subsection{Binary mass ratio}

\begin{figure}
\begin{center}
\centerline{\epsfig{file=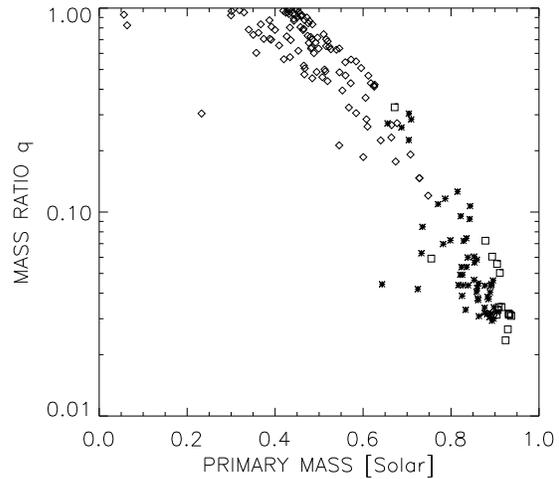,height=7cm}}
\caption{ Mass ratio $q$ versus primary mass, for all the binary and
multiple stars in the simulations. Multiples have been decomposed in
bound two-body systems following the hierarchical configuration (see
text for details). Binaries are represented by diamonds, triples by
asterisks and quadruples by squares. A large fraction of the binaries
have mass ratios in the 0.5-1 range. Only multiples systems can have
very low mass companions in outer wide orbits.}
\end{center}
\end{figure}

Figure~3 depicts the dependence of binary mass ratio $q$ on primary
mass. All systems have been decomposed in pairs, following the
hierarchical distribution. Thus, e.g. a triple will contribute to the
diagram as two binaries: first, the inner two bodies (diamonds), and
second, the pair formed by the centre of mass particle of the inner
binary and the tertiary companion (asterisk). Quadruples are
represented by squares.  Figure~3 illustrates two striking trends that
can readily be related to the processes of gas accretion and dynamical
interactions described in Section~4. Firstly, there is a relative
paucity of binaries with low mass primaries -- essentially there are
no binary primaries with masses less than a factor of 0.5 the minimum
core mass assumed. This is just because the fraction of the core mass
that goes into the single stars/outer multiple companions is very
small and so the binary primary contains a large fraction of the mass
of its parent core. Secondly, there are very few extreme mass ratio
systems and, where they exist, the low mass component is almost always
the wide outermost companion.

The reason for this second trend is rather different from the reason
that extreme mass ratio systems are also rare in pure N-body
simulations. Here, the main effect is not so much that low mass
companions are exchanged out of binaries by more massive ones but
that, instead, once two seeds form a binary, which very often occurs
early on in their accretion histories ($q$ therefore being not far
from unity), the combined gravitational field enhances accretion on to
both components, which subsequently accrete similar amounts of gas.

Likewise, the reason that the binary contains most of the core mass is
not the result of repeated exchange interactions, but instead that,
once two seeds form a binary, they constitute the focus for further
accretion, and thereby outstrip the other cluster members in their
mass acquisition.

Although these results are roughly compatible with the mass ratio
distribution derived by Mazeh et al. (1992) for close G dwarf
binaries, they clearly do not agree with the mass ratio distribution
for G dwarf binaries as a whole: the results of Duquennoy \& Mayor
(1991) show a rise with decreasing mass ratio, at least for mass
ratios as low as $\sim$~0.2.

\begin{figure}
\begin{center}
\centerline{\epsfig{file=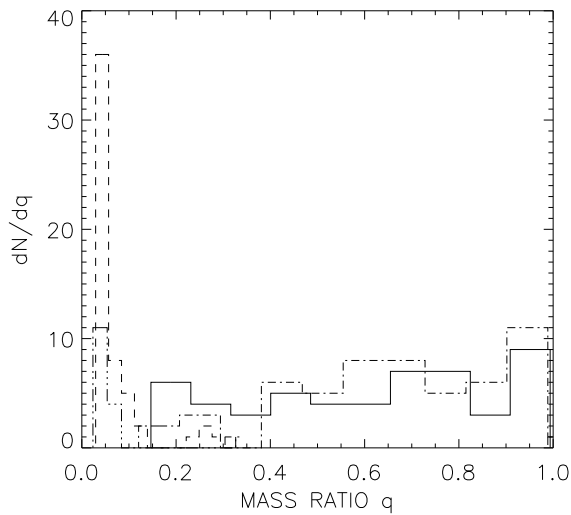,height=7.5cm}}
\caption{ Mass ratio $q$ distribution. The thin solid and dot-dashed line
stand for {\it pure} binaries and inner pairs in multiples,
respectively; the dashed line represent outer pairs in triple systems,
and the double-dot-dashed line outer companions in quadruples. The
binary fraction is an increasing function of $q$, except for
higher-order systems, characterized for extreme values of $q$. Note
how the distribution for {\it pure} binaries is very shallow (flat to
first order), whilst that of inner pairs of multiples increases more
steeply, due to the fact that tertiary companions are in average less
massive than single objects ejected from the cluster -- the reason for
this is that a multiple system can achieve a stable hierarchical
distribution more easily if the outermost companion has a much smaller
mass than each of the binary components.}
\end{center}
\end{figure}

This deficit can be somewhat alleviated if some of the outer pairs in
multiples systems are added to the distribution. As Figure~4 shows,
the rise with increasing mass ratio characteristic of close binaries
(solid and dot-dashed line) contrast with the steep rise in the
distribution at very low mass ratio when all the pairs are included
(e.g. dashed line for triples). This finding is in line with the
general feature of the simulations: low mass companions are much more
likely to be the outlier in triple systems than the secondaries in the
central binary. A prediction of this result is that where wide low
mass companions have been detected, their primaries might, on closer
examination, turn out to be binaries (Gizis et al. 2001).

As mentioned in Section~3.2, the presence of circumstellar discs,
neglected in the present simulations, might contribute to increase the
fraction of low mass ratio systems by acting as a source of energy
dissipation in stellar encounters, or by fragmenting into one
companion when orbiting a single star. Future experiments
(Delgado-Donate et al. in prep.) will focus on larger N ensembles in
order to discover whether ejections due to binary-binary interactions
can give additional insight into the formation of extreme mass ratio
systems. Likewise, the evolution of N=2 systems (studied by Bate 2000
for circular systems), in which competitive accretion is the dominant
process and there is no place for dynamical interactions, will be
explored.

\subsection{Semi-major axis distribution}

\begin{figure}
\begin{center}
\centerline{\epsfig{file=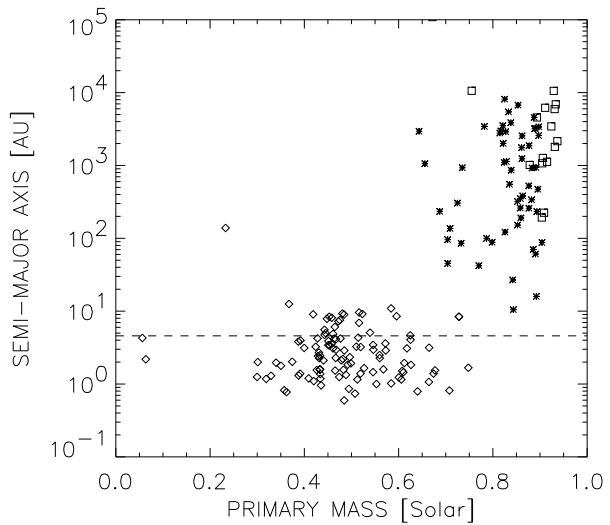,height=7.5cm}}
\caption{Semi-major axis $a$ versus primary mass. Symbol code as in
Figure~3. Mostly close binaries are formed in these simulations: the
mean $a$ is smaller than 10~AU, but follows a Gaussian probability
distribution in logarithmic space (see Duquennoy \& Mayor 1991). This
small $\langle a \rangle$ is due to the initial centrally concentrated
configuration of the stellar {\it seeds} and the number of close
dynamical interactions occurring in the cluster (in turn correlated
with the number of protostars N=5 used in these models). On the
contrary, multiples have large $a$, since the components must follow a
hierarchical distribution, with $a_{\rm outer}$/$a_{\rm inner}$ being
much greater than 10.}
\end{center}
\end{figure}

Figure~5 depicts the semi-major axis dependence on primary mass. As it
can readily be seen, most binaries (diamonds) are rather tight, with a
mean semi-major axis $\langle a \rangle \approx$~5~AU. On the other
hand, wide systems (asterisk for triples, and squares for quadruples)
correspond, with few exceptions, to triple and quadruple stars. This
result can be immediately related to the pattern of dynamical decay
outlined in Section~4.2. Given that a multiple system will only
survive if its orbital elements adopt particular ranges of parameter
space (so that it is sufficiently hierarchical to be dynamically
stable; see e.g. Eggleton \& Kiseleva 1995), triples and quadruples
must display a semi-major axis distribution whose mean is
$\geq$~10~times larger than that of the binaries comprising the inner
components.

The minimum semi-major axis for binaries is set by the numerical
resolution achieved in these calculations, as parameterized by R$_{\rm
min}$. On the other hand, the low mean value of the semi-major axis
distribution is determined by the number and strength of those
dynamical interactions that lead to an ejection. Therefore, $\langle a
\rangle$ is strongly anti-correlated with the number of {\it seeds}
N=5 used throughout these calculations. The initial centrally
concentrated spatial distribution of the protostars (the initial
separation between {\it seeds} is $\sim$~10$^3$~AU, 10~$\times$
smaller than the core radius), is also responsible of the formation of
very close binaries. A more distributed initial configuration would
hace produced a larger fraction of wide binaries.

From these results, together with those described in Section~5.2, it
is clear that there exists a relation between the mass ratio,
semi-major axis and multiplicity of the system. In this kind of
simulations, wide binaries (semi-major axis $>$~10$^3$~AU) could be
formed if there is no orbital decay due to dynamical interactions. For
this to happen the binary components must accrete mass in relative
isolation and therefore, companions at larger separations are not
likely to be present. Since there is no other mechanism that can
reduce the orbit substantially (the collapse of the core induces a
semi-major axis shrinkage of $\sim$ an order of magnitude), the two
stars remain too distant to compete equally for the gas reservoir,
consequently driving the mass ratio quickly to very low values.

\subsection{Eccentricity}

\begin{figure}
\begin{center}
\centerline{\epsfig{file=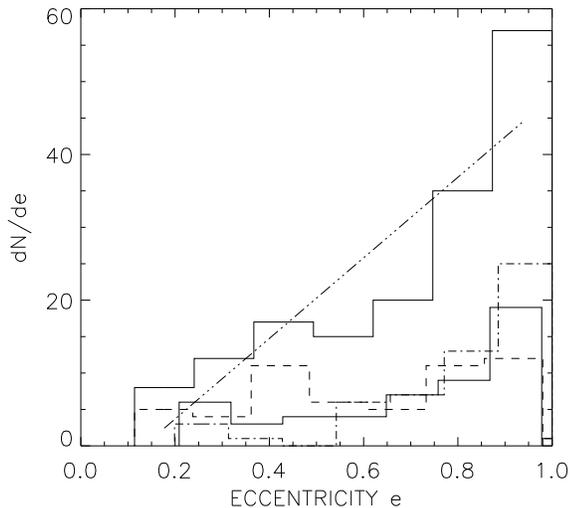,height=7.5cm}}
\caption{ Distribution of orbital eccentricities $e$. The thin solid
and dashed line correspond to double stars and inner pairs of multiple
systems, respectively. The dot-dashed line refers to triple and
quadruple stars. The thick solid line histogram includes all the
pairs. A linear fit is shown to be a reasonable first-order
representation of the total $e$ distribution.}
\end{center}
\end{figure}

Among binary stars (i.e. excluding all outer pairs in multiples), the
eccentricity can take a whole range of values from 0.1 to 1, as shown
by Figure~6. The thick solid line histogram includes all the pairs,
whereas the thin solid line refers to {\it pure} binaries, the dashed
line to the inner binaries in multiples, and the dot-dashed line
correspond to triple and quadruple systems. It is apparent that there
is a preference for high eccentricity values, following a probability
distribution with a functional form f(e)~$\propto$~e. Although the
eccentricity is the orbital parameter most sensitive to star-disc
interactions (high eccentricities would be damped by the accretion of
material from the disc), this result is in good agreement with the
expected distribution of eccentricities for close young binaries (in
which tidal circularization has not had time to operate yet;
Ambartsumian 1937), and long period binaries (Duquennoy \& Mayor
1991).

Multiple systems are characterized by larger eccentricities, as it is
expected for objects which have undergone several close triple
encounters before settling into a stable hierarchical
configuration. This trend is even more accentuated in quadruples than
in triples. Thus, one would expect a multiple system to display an
increasingly larger eccentricity as one moves along the hierarchy, as
well as a progressively smaller mass ratio. This pattern is reflected
in the diagram shown in Figure~6: the eccentricity distribution of
inner binaries (dashed line) is approximately flat whilst that of
outer pairs (dot-dashed line) peaks at very high eccentricities.

\subsection{Kinematics} 

\begin{figure}
\begin{center}
\centerline{\epsfig{file=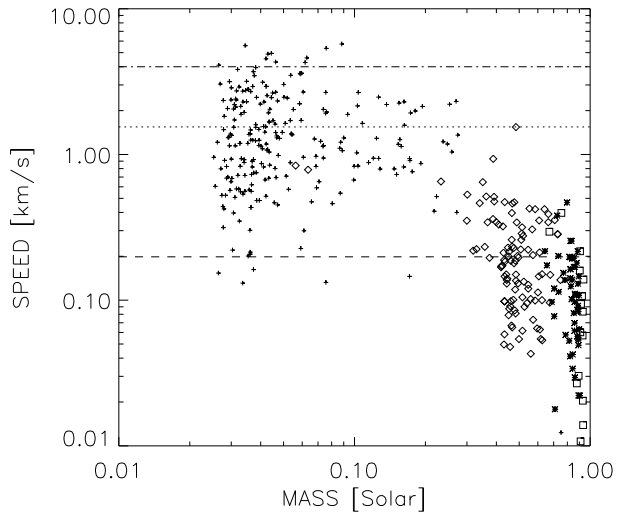,height=7.5cm}}
\caption{ Velocity in km s$^{-1}$ in terms of mass (in M$_\odot$). Crosses
indicate single stars. Binaries and multiples are denoted by the same
symbol code as in Fig~3; for these, the depicted mass is the primary
mass of the pair. The dashed and dotted line indicate the mean
velocity of multiples and singles, respectively. Above the dot-dashed
line are located those objects with a speed larger than 4~km
s$^{-1}$. The velocity offset between multiple systems and singles is
approximatelly one order of magnitude. There is not a clear dependence
of ejection velocity on mass for single objects.}
\end{center}
\end{figure}

The break-up of these small aggregates imprints a clear kinematic
signature on the objects involved. Binary stars attain centre of mass
velocities that, in average, are a factor of 10 smaller than the
typical ejection speed of single stars (see Figure~7; singles are
denoted by crosses and multiples as in previous figures; the dashed
and dotted lines indicate the multiples and singles mean velocity,
respectively). This distinction between the velocity distribution of
singles and binaries is in essence due to the fact that single stars
are all ejected from the core, whereas the binary (or multiple) star
remains close to the core centre of mass. A notable difference between
the N-body case (e.g. Sterzik \& Durisen 1998) and these
hydrodynamical simulations is the relationship between the typical
ejection speeds and the initial parameters of the core. In the
dissipationless calculations, the final velocities of ejected stars
are of the same order as the virial velocity of the initial
cluster. In the simulations reported here there are also a few objects
which are ejected early on with such velocities, where the initial
virial velocity of the core is $\sim$~0.2~km s$^{-1}$. However, most
of the stars are ejected with velocities much greater than this: about
half attain speeds greater than 5 times the initial virial velocity
(the corresponding number for Sterzik \& Durisen is less than 1~\%),
and there is also a significant minority that attain velocities $\geq$
4~km s$^{-1}$ (dot-dashed line in Figure~7). This may be readily
understood inasmuch as in gas dynamical simulations the protostellar
system shrinks as the gas is accreted, so that later interactions are
much closer and produce correspondingly larger velocity kicks.

\begin{figure}
\begin{center}
\centerline{\epsfig{file=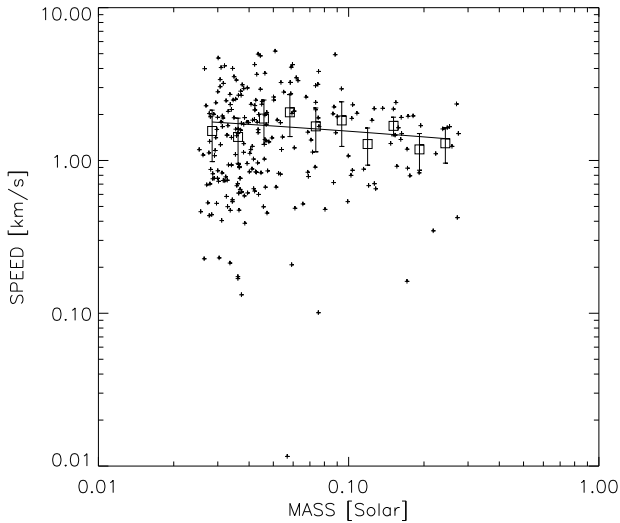,height=7.5cm}}
\caption{ Speed in km s$^{-1}$ versus mass, for single stars only. The
velocities have been averaged in ten logarithmic bins (squares) and
the best linear fit computed. Within the error bars (standard
deviations of the distribution inside every bin), there is no
significant variation of ejection velocity with mass.}
\end{center}
\end{figure}
 
Another noteworthy feature of the kinematics of ejected objects is
shown in Figure~8. The velocity of singles has been averaged in ten
logarithmic bins (squares) spanning the whole range of masses ({\it
per given core mass}). The best fit to these statistical averages does
not display any significant correlation between ejection speed and
escaper mass, within the error bars (given by Poisson noise). This
result arises naturally from the dynamics of the accreting
cluster. Low mass objects can be ejected at all stages of the cluster
evolution since they can remain bound to the cluster at large
distances from the gas-rich centre; thus, their speed pattern will
reflect the different degrees of impulse needed to overcome the
cluster potential as it gets stronger. This {\it modus operandi} is
illustrated by the large spread in velocities shown in Figure~8. Note
that the apparently lower dispersion in ejection velocities for higher
mass singles (mass~$>$~0.06~M$_\odot$) results from the smaller number
of objects located in this mass interval, as demonstrated by a
Kolmogorov-Smirnov test applied to the speed distribution.

\begin{figure}
\begin{center}
\centerline{\epsfig{file=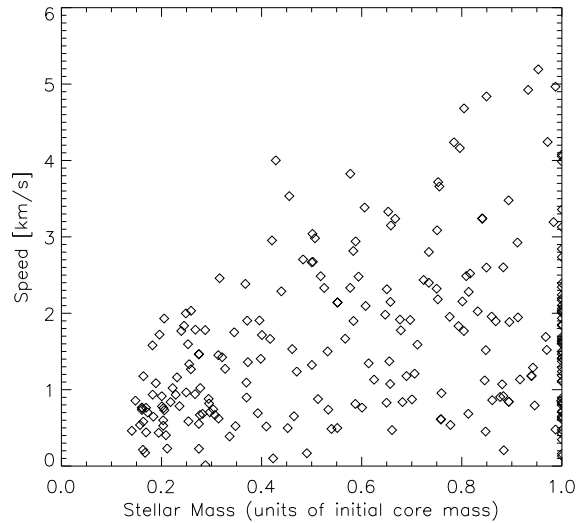,height=7.5cm}}
\caption{ Ejection speed in km s$^{-1}$ versus the mass in the stellar
component at the time of ejection. The {\it stellar mass} is shown in
units of the initial core mass. A clear correlation between these two
quantities can be seen, as it is expected if the distribution of
escaping velocities depends on the binding energy of the cluster at
the time of ejection.}
\end{center}
\end{figure}
Consequently, the escaper's speed depends mainly on the binding energy
of the cluster at the time of ejection (see Bate, Bonnell \& Bromm
2003 for a similar conclusion), which in turn would correlate roughly
with the total mass in the stellar component at that time. This
correlation can be seen in Figure~9.

It is unclear however whether this predicted difference in the
velocity of young single and binary stars would in practice be
detectable in star forming regions, since the velocity offset between
the two (a few km s $^{-1}$) is comparable with the core-core velocity
dispersion in such regions. Additionally, the presence of various
binaries in the same core might spread the velocity dispersion of the
binaries via ejection of some of the pairs, therefore making it less
distinct from the population of escapers (Bate, Bonnell \& Bromm
2003).

\section{The convolution}

In order to relate the findings described in Section~5 to the
resultant IMF, it is necessary to convolve the final stellar masses
{\it per given core mass} (or SMFPD) with a core mass function
(henceforth CMF). This means that the final masses of the protostars
and related quantities are determined by two processes: first, the
{\it seeds} gain mass according to the acquisition pattern found to
operate in these cores; and second, the star {\it obtains} its final
mass once the initial mass of its parent core is determined. In this
operation we are assuming that the division of mass is a scale-free
process. This would certainly be the case either if there were no
systematic dependence on core mass of dimensionless variables such as
e.g. the Jeans number (as controlled by equation (1)), or else if the
results depicted here are insensitive to those variables.

Mathematically, this convolution differs from the usual one in the
sense that two numbers (the stellar mass fraction of a certain {\it
seed} and the mass of the core in which this star is embedded) are
combined by a multiplicative instead of additive operation. That is,
the {\it final} mass of a stellar object m$_*$ will be given by the
product r$_*$~$\times$~M$_c$, where r$_*$ is the mass accreted by the
{\it seed} in units of a given core mass, and M$_c$ is the mass of
that core.

Unfortunately, the choice of a core mass function is not unique since
the process that leads to the formation of dense star-forming cores is
only partially understood. Several functional forms for this CMF are
possible:

\begin{enumerate}
\item A power-law with index of $-$2, as expected from the equipartition
of bulk kinetic energy and gravitational energy in a pressureless
fluid.

\item A log-normal mass function, as derived from the density
probability distribution function of Jeans unstable clumps in computer
simulations of compressible magnetohydrodynamic turbulence (Padoan et
al 1997; Klessen 2001).

\item A power-law with Salpeter or nearly-Salpeter slope
($\alpha$~=~$-$2.35), as found by Motte et al. (1998) in the mass
distribution of dense cores in Ophiucus.

\item A combination of power-law and log-normal distributions, as
proposed by Padoan \& Nordlund (2002) based on turbulent fragmentation
calculations.
\end{enumerate}

In order to fully understand the effects that a {\it multiplicative}
convolution has on the CMF, we must note that a SMFPD described by a
delta function would conserve the shape and slope of the CMF, and
simply displace it to the position of the delta. If instead of a delta
function we use a narrow box function, the CMF is additionally
stretched to the new limits given by [~r$_{\rm
*,min}$~$\times$~M$_{\rm c,min}$~,~r$_{\rm *,max}$~$\times$~M$_{\rm
c,max}$~]. Also, the convolution of two power-laws -- e.g. a SMFPD of
slope $- \alpha$ confined between 0 and 1, and a CMF of slope $-
\beta$ with an arbitrary larger upper cutoff (where $\alpha$ is
assumed to be smaller than $\beta$) -- would, in general, give rise to
a new function characterized by a break at a position given by the
product of the lowest CMF value and the highest SMFPD value. This
break would split the new function into two regions: at lower values,
a power-law of slope $- \alpha$, and at higher values, a power-law of
slope $- \beta$. Note that $\alpha$ and $\beta$ would exchange their
range of influence if we choose $\alpha$ to be larger than $\beta$;
that is, the steepest slope dominates at larger values.

\subsection{Initial Mass Function}

\begin{figure}
\begin{center}
\centerline{\epsfig{file=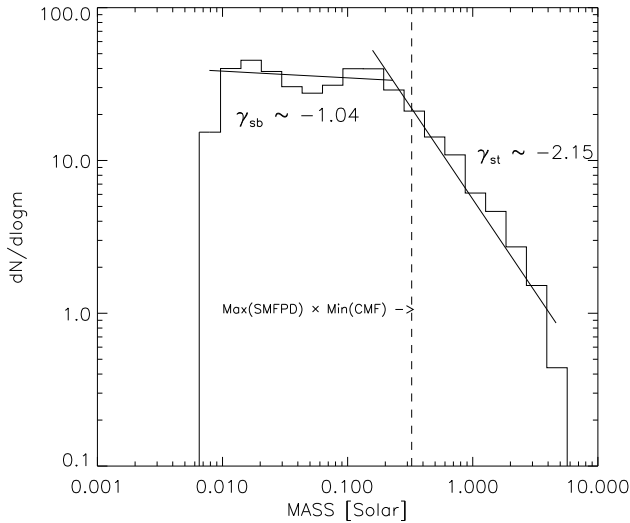,height=7.5cm}}
\caption{ Initial Mass Function (IMF) for the stars and brown dwarfs
formed in the simulations. This IMF is the result of the convolution
of the bimodal SMFPD with a core mass function (CMF) taken to be a
power-law of slope $\alpha$=$-$2.35 (Salpeter) and cutoff at 0.25 and
10~M$_\odot$. For masses approximately above the product of the
largest SMFPD value with the lowest CMF value, the mass function
resembles the observed IMF. Below, the distribution is approximately
flat in logarithmic scale (slope~$\approx$~$-$1).}
\end{center}
\end{figure}

Figure~10 shows the result of a convolution of the bimodal SMFPD with
a CMF taken to be a power-law of index $-$2.35 (Salpeter IMF) in the
range of core masses 0.25 to 10 M$_\odot$. The choice of the power
index and upper cutoffs are motivated by the millimetre continuum
observations of Motte et al. (1998) and the H$^{13}$CO$^+$ survey of
Onishi et al. (2002), respectively. The lower cutoff comes from the
minimum mass a core can have so that the initial {\it seed} mass lies
above the opacity limit for fragmentation ($\sim$~5 Jupiter masses;
Low \& Lynden-Bell 1976; Rees 1976; Boss 1989; Bate 1998, 2002), given
the initial conditions imposed -- 10\% of the mass is in stellar form
at the start of the calculations. Clearly, the resulting IMF
essentially follows the core mass function down to a stellar mass
close to the minimum core mass. The reason for this correspondence is
that for a steep core mass function, the majority of stars of a given
mass belong to the {\it high mass peak} of low mass cores, rather than
the {\it low mass peak} of the less numerous higher mass cores. Thus
the number of stars of given mass is governed by the frequency of
cores whose mass exceeds the stellar mass by a rather modest factor
(less than a factor two) and the IMF therefore follows the core mass
function.

Similarly, convolutions performed using log-normal mass functions as
CMF, in which the characteristic mass and standard deviation adopt
typical values, and the cutoffs are set at the same values chosen for
the power-law CMF discussed above, produce invariably the same
outcome. That is, the resulting IMF resembles the original log-normal
mass function and simply differs in its width, which stretches to the
values set by the products of minimum and maximum core and stellar
masses. Also, as it may be expected from the previous results, the
functional form proposed by Padoan \& Nordlund (2002) for the CMF (in
the original, IMF) is equally preserved.

Thus we find that although star formation in this case is far from
being a one to one mapping between core and star mass, the IMF of the
stars is controlled by, and nearly identical to, the core mass
function. We note that the observed similarity between the IMF and
core mass function {\it cannot} therefore be used as an argument
against multiple fragmentation and competitive accretion inside
star-forming cores.

\subsection{Binary Properties}

The convolution of core masses with a CMF has an effect on the
resulting properties of binary stars. Although the binary mass ratio
and eccentricity is independent of the individual masses of the binary
components, the dependence of these parameters on the mass of the
primary is modified by the CMF assumed. Additionally, if the initial
mass of a core and its radius are assumed to be correlated in some way
(as it is the case when e.g. the initial cloud conditions are set
according to equation (1)), the semi-major axis distribution is also
affected by the choice of the CMF. Equation (1) also states that the
final velocity dispersion is independent of the initial mass assumed
for each individual core.

\begin{figure}
\begin{center}
\centerline{\epsfig{file=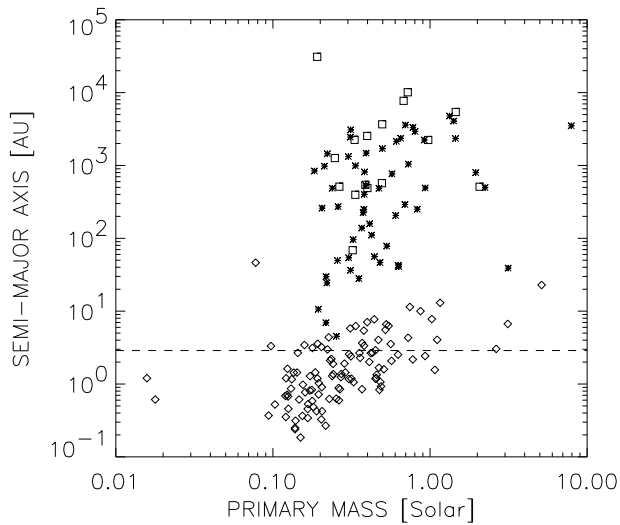,height=7.5cm}}
\caption{Same as in Figure~5, after performing a
convolution with a power-law CMF.}
\end{center}
\end{figure}

Assuming (1) to hold, and making use of the same CMF as in
Section~6.1, we have derived the scaled primary masses and semi-major
axis of all the pairs. Fig~11 shows the semi-major axis dependence on
primary mass. The linear relation between primary mass and semi-major
axis imposed by equation (1) is reflected in the diagram, as well as
the effect of the convolution, which has smeared out the distribution
of points initially clustered around the original mean value. From
this diagram, it can also be seen that the process of convolution with
a CMF results in a large fraction of binaries with low mass primaries
-- binary primaries have a mass $\sim$~0.5~$\times$ the core mass
assumed, but there are many more cores with masses below 1~M$_\odot$
than above.

\begin{figure}
\begin{center}
\centerline{\epsfig{file=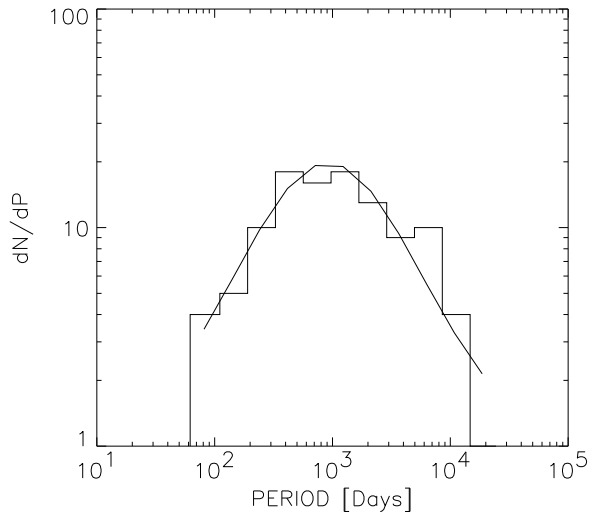,height=7.5cm}}
\caption{Distribution of orbital periods for the binary stars produced
in these simulations. Shown is a Gaussian fit to the histogram.}
\end{center}
\end{figure}

The histogram of periods is shown in Fig~12. It can be fitted by an
approximate Gaussian distribution which peaks at $\sim$~10$^3$~days.
Patience et al. (2000) find a similar value for binaries in dense
environments (open clusters), whereas Duquennoy \& Mayor (1991) show
an observational maximum at $\sim$~10$^4$~days for field G type stars
(comparable to that of T Tauri stars).  This discrepancy might be due
to the number of stellar objects imposed as initial conditions in the
model. Systems where repeated strong interactions are not so likely
might produce longer period binaries -- for this to be a solution to
the lack of wide binaries in these simulations, the fraction of cores
with two initial seeds must be comparable or larger than that of cores
possesing more substructure. The addition to the gas of some initial
angular momentum might as well enhance the fraction of binaries with
longer periods.

In summary, the choice of an observationally-based CMF results in a
set of binary masses and orbital parameters that are in reasonable
agreement with observations. The formation of wide binaries, as
suggested above, might be accomplished in a scenario where dynamical
interactions were not important, as in cores where only two or three
collapsed fragments are present. The origin of very low mass ratio
binaries, however, remains obscure: both pure N-body calculations
(e.g. Sterzik \& Durisen 1998) and the present hydrodynamic-N-body
models fail to produce as large a fraction of extreme mass ratio
binaries as observational data suggest (the Duquennoy \& Mayor mass
ratio distribution peaks at a value of 0.2). Although some of these
double stars might turn out to have internal substructure, in the
sense of the inner body being in fact a binary star (e.g. Tokovinin \&
Smekhov 2002), -- this would be immediately accounted for in our
simulations --, it is obvious that very low mass ratio binaries
constitute a significant population and its origin needs to be further
investigated.

\section{Brown Dwarfs}

Once a convolution with a suitable CMF is performed (e.g. a power-law
with nearly Salpeter slope within the observed mass range), the
calculations show that the population of escapers and outer companions
is abundant in brown dwarfs.

As described in Section~4.2, low mass objects are naturally produced
when dynamical interactions within an unstable accreting multiple are
taken into account. These interactions provide a simple dynamical
explanation for the formation of sub-stellar objects, which start with
a mass of the order of the opacity limit of fragmentation and are
prevented from reaching the hydrogen burning limit due to the limited
amount of gas they can accrete before they are ejected from the common
envelope (Reipurth \& Clarke 2001; Bate et al 2002a).

\begin{figure}
\begin{center}
\centerline{\epsfig{file=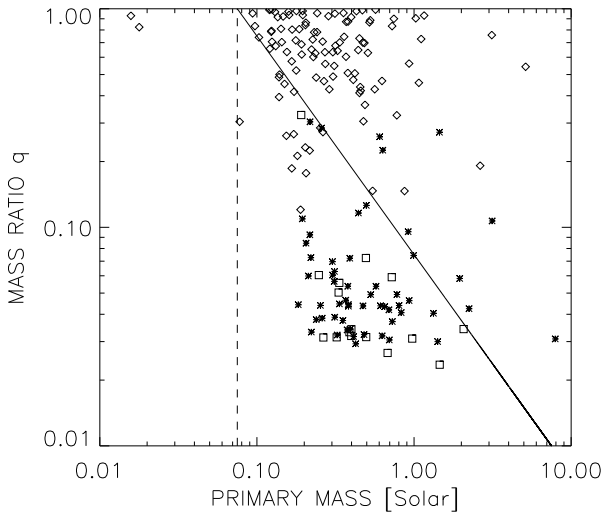,height=7.5cm}}
\caption{Same as in Figure~3, after applying a convolution with
a CMF as described in text. Note that here both axis are in
logarithmic scale.}
\end{center}
\end{figure}

Figure~13 shows the mass ratio of multiple systems in terms of the
mass of the primary component. As explained in Section~5.2, all
multiples have been partitioned in two-body systems (symbol code as in
previous figures). Below the solid line are located those pairs which
posses a brown dwarf companion; binary brown dwarfs occupy the region
to the left of the dashed line. It can be readily seen that brown
dwarfs {\it produced by} dynamical interactions are infrequent as
binary companions. On the other hand, among hierarchical triples and
quadruples the outermost object is very often a brown dwarf, with a
probability higher than 90 \% for low-mass primaries -- this fraction
is certainly an upper limit, since some of the triple and quadruple
systems included in this study might not be stable enough to survive
the pre-main-sequence stage. In this sense, no brown dwarf desert at
wide separations should be apparent if the inner components were not
resolved (see Gizis et al. 2001). Brown dwarfs can be wide binary
companions, but in this case the {\it primary} is predicted either to
be a binary system or to have a low mass (less than
$\approx$~0.3~M$_\odot$). Binary brown dwarfs are equally improbable
in this scenario, with the exception of some scarce, tight, nearly
equal mass pairs.

The above results on brown dwarf binary pairings represent a critical
area for confronting dynamical decay models with observations. It is
currently unclear how the binary fraction changes in the case of very
low mass primaries, since the detection of brown dwarf binaries is in
its infancy (see, for example, Reid et al. 2001). If the binary
fraction were indeed lower for brown dwarfs, it could readily be
explained by such models, whereas a high incidence of brown dwarf
binaries would be explicable only if the core mass function extended
to sufficiently low masses (i.e. close to or below the hydrogen
burning mass limit). Thus the detection of a large population of
binary brown dwarfs would imply that most brown dwarfs arose as the
majority component of low mass cores, rather than being the low mass
ejectae from much more massive cores. Clearly, a firmer estimate of
the brown dwarf binary fraction would place important constraints on
formation models.

It is also worth noting that, in these models with minimum core mass
of 0.25~M$_\odot$, the fraction of brown dwarfs that remain in bound
systems is approximately 25\% (75\% are ejected as singles). This is
in clear contrast to objects above the hydrogen burning limit, which
display a {\it boundedness} fraction close to 90\%.

\begin{figure}
\begin{center}
\centerline{\epsfig{file=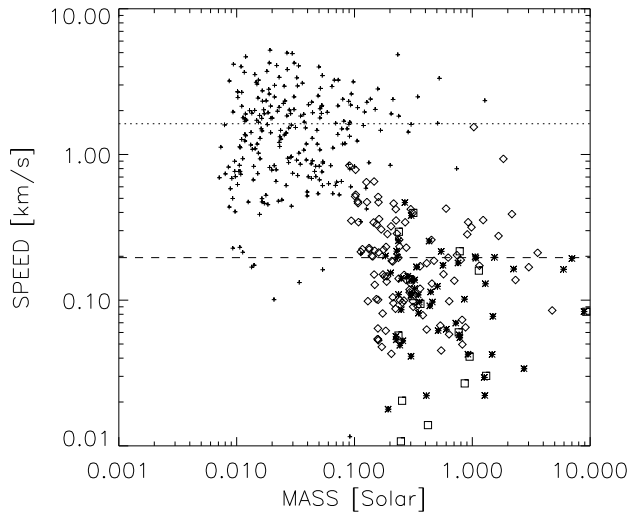,height=7.5cm}}
\caption{Same as in Figure~7. The masses have been
convolved using a CMF as described in the text.}
\end{center}
\end{figure}

The kinematics of sub-stellar objects are also likely to reflect their
origin.  Although there is no appreciable dependence of the final
velocity of the singles on mass, brown dwarfs acquire as a class
higher velocities than stellar objects, including binary and multiple
systems (see Figure~14; symbols and lines as in Figure~7). With
velocities of a few km s$^{-1}$, it is possible that ejected brown
dwarfs would be detectable in the vicinity of Class O objects, as
suggested by Reipurth \& Clarke (2001). Relatively few brown dwarfs
have velocities in excess of $5$ km s$^{-1}$, however, so that the
bulk of brown dwarfs in the Pleiades would have been retained, given
that the escape velocity in the young Pleiades is likely to have
exceeded this value. On the other hand, smaller stellar associations
such as Taurus-Auriga could have already lost a significant fraction
of its sub-stellar population and therefore, would show a relative
excess of binary stars. The findings of Brice\~{n}o et al. (2002) are
in line with this argument: their survey indicates that Taurus has
$\sim$~2~$\times$ fewer brown dwarfs at 0.02-0.08~M$_\odot$ than the
Trapezium, whereas it is well known (Ghez et al. 1993; Leinert et
al. 1993; Simon et al. 1995, K\"{o}hler \& Leinert 1998) that the
binary frequency among T Tauri stars in Taurus is enhanced by a factor
of two compared to solar type stars on the main sequence.

\section{Conclusions}

We have modelled the decay of non-hierarchical gas rich systems
comprising 5 seed protostars in order to study the mass function,
binarity and kinematics of the resulting population. This study can be
seen as complementary to that of Bate, Bonnell
\& Bromm (2002a,b; 2003) who have modelled similar processes in the context
of the calculation of the global evolution of a molecular
cloud. Although the approach used here has the disadvantage that its
initial conditions are artificial, and that it evidently cannot treat
any possible interactions between star forming cores, the
computational economy effected in not treating the whole cloud ensures
a) that one can obtain statistically significant populations of stars
for a given initial condition and hence isolate the factors that
determine the properties of the resulting stars and b) that one can
pursue the evolution to {\it completion}, i.e. to the point where all
the gas has been accreted and where the system has decayed into a
hierarchical multiple system. This latter feature allows us, in
contrast to the calculations of Bate et al., to explore the properties
of wider binaries, which remain undecayed in the Bate et al
simulations.

Our chief conclusions are as follows:
\begin{enumerate}
\item The mass function of the resulting stars is primarily determined
by the mass function of parent cores, being comparable to this core
mass function down to stellar masses that are similar to the minimum
core mass, and declining at masses below this. The reason for this is
that the division of mass {\it for given core mass} is essentially
bimodal - there is a high mass peak corresponding to a binary pair and
a lower mass peak corresponding to ejected singles or wide tertiary
companions. When this is convolved with a core mass function that
declines steeply towards high masses, the majority of stars of given
mass belong to the high mass peak of low mass cores rather than the
low mass peak of rare, high mass cores. Consequently, the stellar mass
function follows the core mass function. Thus the observed similarity
between the core mass function and the stellar IMF (Motte et al. 1998)
{\it cannot of itself be used to disprove the hypothesis that most
stars arise in small non-hierarchical multiples}.
\item Stars much less massive than the  minimum core mass are
unlikely to be binary primaries. This can be understood in terms of
the evolutionary pattern described above. Thus if a large fraction of
brown dwarfs turn out to be binary primaries, this must imply that the
parent core distribution extends well into the substellar regime. In
this case, although brown dwarfs would form in non-hierarchical
multiples, the brown dwarf binaries would not be ejected from these
systems but would represent the highest mass objects in their parent
cores. {\it Thus a firm evaluation of the binary fraction amongst
brown dwarfs will place important constraints on the properties of
their parent cores}. [There is the additional possibility that
star-disc interactions can produce tight brown dwarf-brown dwarf
binaries that would survive ejection. The lack of discs in the
simulations reported here do not allow us to model this process. The
simulations of Bate et al. (which do include discs) indicate that this
is not common however].
\item Binary pairs with extreme mass ratios ($q < 0.2$) are
not produced in the simulations. More extreme mass ratios only occur
in the case of low mass tertiary companions in a wide orbit around a
central tight binary. Such a configuration is the natural consequence
of the decay of non-hierarchical systems, in which low mass stars may
either be ejected or else end up in wide hierarchical orbits.  We
point out that it is an expectation of dynamical decay models that
{\it where brown dwarfs are found as distant companions to normal
stars} (Gizis et al 2001; Kirkpatrick et al. 2001), {\it the central
object should normally be a close binary}. Over all, the distribution
of binary mass ratios (Figure 4) is well matched to the observations
of Mazeh et al (1992) for close G dwarf pairs but is probably
deficient in low mass companions when compared with the statistics of
G dwarf binaries as a whole (Duquennoy and Mayor 1991).
\item Single stars are ejected with typical velocities of a few km s$^{-1}$
(Figure 7) irrespective of their mass. The final velocities of binary
stars with respect to their parent cores is typically an order of
magnitude lower. {\it In principle, this signature of dynamical decay
could be sought in star forming regions}, although may be masked by
the relative motion of the parent cores (typically at a few km
s$^{-1}$). Since the binary fraction is lower among low mass stars and
brown dwarfs than in higher mass stars (see (ii) above), this
depedence of ejection speed on binarity means that in practice brown
dwarfs are born with a higher velocity dispersion (with respect to
their parent cores) than higher mass stars.  {\it With typical
velocities of a few km s$^{-1}$, the majority of brown dwarfs would be
expected to be retained in open clusters}.
\end{enumerate}

\section*{Acknowledgments}

We thank the referee for comments that helped to improve the paper,
and Jim Pringle, Cristiano Porciani and Bo Reipurth for their useful
input. EJDD is grateful for support to the European Union Research
Training Network {\it The Formation and Evolution of Young Stellar
Clusters}.

\label{lastpage}

\end{document}